\documentclass[prd,twocolumn,showpacs,superscriptaddress]{revtex4}

\usepackage{amsmath,amssymb}
\usepackage{verbatim}
\usepackage{graphicx}
\usepackage{hyperref}
\usepackage{color}

\DeclareFontFamily{OT1}{rsfs}{}
\DeclareFontShape{OT1}{rsfs}{m}{n}{ <-7> rsfs5 <7-10> rsfs7 <10->rsfs10}{} 
\DeclareMathAlphabet{\mycal}{OT1}{rsfs}{m}{n}

\newcommand{\be}[1]{ \begin{equation}\label{#1} }
\newcommand{\ee}{\end{equation}}
\newcommand{\bea}[1]{\begin{eqnarray}\label{#1} }
\newcommand{\eea}{\end{eqnarray}}

\newcommand{\eq}[2]{\begin{equation} #1 \label{#2} \end{equation}}

\newcommand{\de}{\delta}
\newcommand{\om}{\omega}
\newcommand{\ka}{\kappa}

\DeclareMathOperator{\extdm}{d}
\newcommand{\extd}{\extdm \!}

\begin{document}

\title{Anti-de~Sitter holography for gravity and higher spin theories in two dimensions}

\author{Daniel Grumiller}
\email{grumil@hep.itp.tuwien.ac.at}
\affiliation{Institute for Theoretical Physics, Vienna University of Technology, Wiedner Hauptstrasse 8--10/136, A-1040 Vienna, Austria}

\author{Mauricio Leston}
\email{mauricio@iafe.uba.ar}
\affiliation{Instituto de Astronomia y Fisica del Espacio (IAFE), Casilla de Correo 67, Sucursal 28, 1428 Buenos
Aires, Argentina}

\author{Dmitri Vassilevich}
\email{dvassil@gmail.com}
\affiliation{CMCC, Universidade Federal do ABC, Santo Andr\'e, 0910-580, S.P., Brazil}

\date{\today}

\preprint{TUW--13--18}

\begin{abstract} 
We provide a holographic description of two-dimensional dilaton gravity with Anti-de~Sitter boundary conditions. 
We find that the asymptotic symmetry algebra consists of a single copy of the Virasoro algebra with non-vanishing central charge and point out difficulties with the standard canonical treatment. 
We generalize our results to higher spin theories and thus provide the first examples of two-dimensional higher spin gravity with holographic description. For spin-3 gravity we find that the asymptotic symmetry algebra is a single copy of the W$_3$-algebra.
\end{abstract}


\maketitle

\section{Introduction}

Gravity models in lower dimensions can provide useful insights into classical and quantum gravity. They were studied vigorously in the past three decades. In terms of technical simplicity, the optimal choice for the dimension is two: it is the lowest dimension where the notions of curvature, causal structure, light-cones and black holes exist, all of which are essential features of the way we think about gravity in higher dimensions. Of course, some aspects of higher-dimensional gravity are inevitably lost --- for instance, there are no graviton excitations in two dimensions (2d) --- but if these aspects are not of relevance for a given physical question then studying 2d toy models can be a rewarding exercise.

The first thing to realize when working in 2d is that Einstein gravity is not the right starting point: any 2d metric trivially solves the 2d Einstein equations as a consequence of the 2d identity $R_{\mu\nu}=\tfrac12\,R g_{\mu\nu}$. The most suitable set of theories are scalar-tensor theories, also known as dilaton gravity. These theories have non-trivial equations of motion (EOM) and non-trivial solutions, including black holes. 

Indeed, 2d dilaton gravity has been employed to study, among other things, black holes in string theory \cite{Mandal:1991tz,Elitzur:1991cb,Witten:1991yr}, black hole evaporation \cite{Callan:1992rs}, black hole complementarity \cite{Susskind:1993if}, black hole thermodynamics \cite{Gegenberg:1994pv,Grumiller:2007ju}, information loss \cite{Fiola:1994ir,Ashtekar:2008jd}, the S-matrix in quantum gravity \cite{Fischer:2001vz}, and gravity at large \cite{Grumiller:2010bz} and small \cite{Carlip:2011uc} distances. See \cite{Grumiller:2002nm} and references therein for further literature on 2d dilaton gravity. 

An item conspicuously absent in this list is holography \cite{'tHooft:1993gx,Susskind:1995vu} and the Anti-de~Sitter/Conformal Field Theory (AdS/CFT) correspondence \cite{Maldacena:1997re}. The reason for its absence is because so far no satisfying treatment of holography exists in 2d dilaton gravity, despite of several interesting attempts like \cite{Strominger:1998yg,Cadoni:1999ja,NavarroSalas:1999up,Hartman:2008dq}. We shall comment on them (and on further work) as we go along. 

Given the recent excitement about higher spin holography in three dimensions \cite{Henneaux:2010xg,Campoleoni:2010zq,Gaberdiel:2010pz,Afshar:2013vka,Gonzalez:2013oaa} another important item that is missing on the list above is the construction of 2d higher spin theories and their holographic description. 
These two items are the main motivation for our work.

The main purpose of this paper is to establish AdS holography in 2d dilaton gravity and for the first time also in higher spin theories in 2d.

The 2d dilaton gravity bulk action \cite{Banks:1991mk,Russo:1992yg,Odintsov:1991qu}
\begin{equation}
S=\frac{\ka}{2\pi}\, \int\extd^2x \sqrt{g}\, \big(X R + U(X)\,(\partial X)^2 + 2V(X) \big)\label{HS}
\end{equation}
contains the 2d gravitational coupling $\ka$, the dilaton field $X$ and two arbitrary potentials thereof, $U(X)$ and $V(X)$. We set $\ka=1$, work in Euclidean signature throughout (though most of our results extend straightforwardly to Lorentzian signature) and restrict for the time being to models with $U(X)=0$.

Dilaton gravity in 2d is locally quantum trivial \cite{Kummer:1996hy}, but globally it can be non-trivial, which is why physical boundary states could exist, similar to the situation in three-dimensional Einstein gravity with negative cosmological constant \cite{Brown:1986nw} or flat space chiral gravity \cite{Bagchi:2012yk}. It is one of the aims of this paper to check to what extent this is true.

\section{Preliminaries}

Like in three-dimensional gravity, where a gauge-theoretic formulation as Chern--Simons theory exists \cite{Achucarro:1986vz,Witten:1988hc}, there is a useful gauge-theoretic formulation of 2d dilaton gravity as a non-linear gauge theory \cite{Ikeda:1993fh}, namely a Poisson-sigma model (PSM) \cite{Schaller:1994es} (see \cite{Strobl:2004im} for higher-dimensional generalizations).
\begin{multline}
S=\frac 1{\pi}\, \int\extd^2x\, \tilde\epsilon^{\mu\nu}\, \big( X^a (\partial_\mu e_{\nu a}+\omega_\mu \epsilon_{a}{}^{b}e_{\nu b}) +X\partial_\mu\omega_\nu \\
+\tfrac 12 \epsilon^{ab}e_{\mu a}e_{\nu b} V(X) \big) \label{1stor}
\end{multline}
The notations mostly follow \cite{Grumiller:2002nm} (except for our Euclidean signature). Latin indices are raised and lowered with $\de_{ab}={\rm diag}(1,\, 1)_{ab}$.  We set $\epsilon^{10}=1$. We denote the ``holographic'' (or radial) coordinate by $\rho$ and the angular coordinate (or in Lorentzian signature: time coordinate) by $\varphi$, identifying $\varphi\sim\varphi+2\pi$. We fix the sign in the Levi-Civita symbol $\tilde\epsilon^{\mu\nu}$ by $\tilde\epsilon^{\rho\varphi}=1$.

To make the relation of the first order action \eqref{1stor} to a PSM manifest we rewrite it as
\begin{equation}
S=\frac 1{\pi}\, \int\extd^2x\, \tilde\epsilon^{\mu\nu}\, \big( X^I\partial_\mu A_{\nu I} +\tfrac 12 P^{IJ}(X^K) A_{\mu I}A_{\nu J}\big)\label{PSM}
\end{equation}
with three target space coordinates $X^I=(X, X^a)$, three connection 1-forms $A_x=\om$, $A_a=e_a$, and the Poisson tensor ($P^{IJ}=-P^{JI}$)
\begin{equation}
P^{Xb} = X^a\epsilon_{a}^{\ b}\,,\qquad P^{ab} = V(X)\epsilon^{ab} \,.
\label{Ptensor}
\end{equation}
As a consequence of the non-linear Jacobi identities
\eq{
\frac{\partial P^{IJ}}{\partial X^L}\,P^{LK} + \frac{\partial P^{JK}}{\partial X^L}\,P^{LI} + \frac{\partial P^{KI}}{\partial X^L}\,P^{LJ} = 0
}{eq:2d1}
the non-linear gauge transformations
\begin{subequations}
\label{PSMgauge}
\begin{eqnarray}
&&\delta_\lambda X^I=P^{IJ}\lambda_J\\
&&\delta_\lambda A_{\mu I}=-\partial_\mu\lambda_I -\frac{\partial P^{JK}}{\partial X^I} \lambda_K A_{\mu J}
\end{eqnarray}
\end{subequations}
leave the PSM action (\ref{PSM}) invariant up to a total derivative,
\begin{equation}
\delta_\lambda S=\frac 1{\pi}\, \int\extd^2x\, \partial_\mu \Big[ \tilde\epsilon^{\mu\nu} A_{\nu I}\lambda_J \Big( P^{IJ} -X^K \frac{\partial P^{IJ}}{\partial X^K} \Big) \Big]\,.\label{totder}
\end{equation}
Terms in $P^{IJ}$ that are linear in $X^K$ do not contribute to the gauge variation (\ref{totder}).

In components the gauge transformations \eqref{PSMgauge} read
\begin{subequations}
 \label{eq:2d6}
\begin{eqnarray}
&&\delta_\lambda X^a = V\epsilon^ {ab}\lambda_b -X^b\epsilon_b^{\ a}\lambda_X\\
&&\delta_\lambda X=X^b\epsilon_b^{\ a}\lambda_a\\
&&\delta_\lambda e_{\mu a}= -\partial_\mu \lambda_a -\omega_\mu \epsilon_a^{\ b}\lambda_b
+\epsilon_a^{\ b} e_{\mu b}\lambda_X\\
&&\delta_\lambda \omega_\mu = -\partial_\mu \lambda_X - \epsilon^{ab} e_{\mu a}\lambda_b\,\extd V / \extd X
\end{eqnarray}
\end{subequations}

Canonically, the gauge transformations \eqref{eq:2d6} (which correspond on-shell to diffeomorphisms and Lorentz transformations) are realized through first class constraints. 
\begin{equation}
\partial_\rho X^I +P^{IJ}A_{\rho J}=0\label{cons}
\end{equation}
In components the constraints read as follows.
\begin{subequations}
\label{eq:2d7}
\begin{eqnarray}
&&\partial_\rho X^0 - X^1\omega_\rho - Ve_{\rho 1}=0 \label{cons0}\\
&&\partial_\rho X^1 + X^0\omega_\rho + Ve_{\rho 0}=0 \label{cons1}\\
&&\partial_\rho X + X^1e_{\rho 0} - X^0e_{\rho 1}=0 \label{consX}
\end{eqnarray}
\end{subequations}
The remaining field equations are the torsion constraint
\eq{
\tilde\epsilon^{\mu\nu}\,\big( \partial_\mu e_{\nu a}+\omega_\mu \epsilon_{a}{}^{b}e_{\nu b}\big)=0\,,
}{eq:2d42}
which allows to express the spin-connection $\om_\mu$ through the zweibein $e_\mu^a$, the curvature equation
\eq{
R = \frac{2\tilde\epsilon^{\mu\nu}}{\det(e)}\,\partial_\mu\omega_\nu = -2\,\extd V/\extd X
}{eq:angelinajolie}
and the same equations as the constraints \eqref{eq:2d7}, but with $\rho$ replaced by $\varphi$.

There are two distinct sets of solutions to the field equations. Constant dilaton vacua, with $X^0=X^1=V(X)=0$ and $X=\rm const$., and linear dilaton vacua, where the dilaton $X$ does depend on $\rho$. The latter solutions are generic, while the former exist only for specific models, and even then require an infinite finetuning for the value of the dilaton. Some recent attempts towards AdS holography for constant dilaton vacua are based on work by Hartman and Strominger \cite{Hartman:2008dq,Castro:prep} and require coupling to a Maxwell field. Also the recent construction of spin-3 gravity by Alkalaev restricts to the constant dilaton sector \cite{Alkalaev:2013fsa}. 

We do not consider this non-generic sector and focus instead on generic linear dilaton vacua. At first glance it may seem surprising that asymptotic AdS symmetries can be compatible with linear dilaton vacua, since a vector field $\xi$ that generates diffeomorphisms does only preserve a constant dilaton, $\xi^\mu\partial_\mu X = 0$. However, it is sufficient if these vector fields preserve the asymptotic structure of the dilaton (and of all other fields involved), and this depends on the precise boundary conditions one imposes, which turn out to be somewhat delicate. It is probably for this reason that so far no consistent holographic description exists for linear dilaton vacua.

\section{AdS$_2$ boundary conditions}

The discussion above applies to arbitrary potentials $V(X)$. However, in order to obtain asymptotically AdS solutions (with unit AdS radius) the potential must asymptote to $X$, since only then the Ricci scalar asymptotes to $R=-2$, see \eqref{eq:angelinajolie}. For simplicity let us therefore consider from now on the potential $V(X)=X$, which describes the Jackiw--Teitelboim model \cite{Jackiw:1984,Teitelboim:1984} and postpone a description of generalizations thereof.
In that case the PSM reduces to a linear gauge theory, with gauge group $SL(2)$ \cite{Isler:1989hq,Chamseddine:1989yz,Cangemi:1992bj,Achucarro:1992mb}.

We discuss now locally asymptotically (Euclidean) AdS boundary conditions. We assume that $\rho\to\infty$ corresponds to the right (R) AdS boundary and $\rho\to-\infty$ to the left (L) one. For simplicity we choose the gauge
\eq{
e^1_\rho = 1 \qquad e_\rho^0 = e^1_\varphi = 0\,.
}{eq:2d2}
The only non-trivial zweibein-component is $e^0_\varphi$.
It must diverge exponentially in $\rho$ at {\em both} AdS boundaries.
Thus, a reasonable Ansatz is
\eq{
e^0_\varphi = T_R(\varphi)\,e^\rho + T_L(\varphi)\,e^{-\rho}
}{eq:2d3}
where $T_{R, \,L}$ are arbitrary (state-dependent) functions of the angular coordinate.
We note that the exponential behavior in $\rho$ of the zweibein generalizes one of the assumptions imposed in a previous approach to PSM holography by one of us \cite{Vassilevich:2013ai}. 

In fact, solving the EOM in the gauge \eqref{eq:2d2} it can be shown that \eqref{eq:2d3} is the most general solution, which shows the consistency of our Ansatz \eqref{eq:2d3}. In the language of Brown and Henneaux \cite{Brown:1986nw} we have gauge-fixed small gauge transformations by the choice \eqref{eq:2d2} and parametrized the large gauge transformations (those that change the physical state of the theory) by the functions $T_{R, \,L}(\varphi)$.

With the gauge \eqref{eq:2d2} and boundary conditions \eqref{eq:2d3} on the zweibein we proceed now to determine the boundary conditions for the connection by demanding consistency with the torsion constraint \eqref{eq:2d42}.
\begin{equation}
\omega_\varphi=-T_R(\varphi)\, e^\rho + T_L(\varphi)\, e^{-\rho} \qquad \omega_\rho=0 \label{omegas}
\end{equation}

We demand that the target space coordinates obey boundary conditions such that the constraints \eqref{eq:2d7} hold identically; this means nothing else but prohibiting boundary conditions that violate the constraints and is therefore a meaningful restriction. 
\eq{
X^0 = \partial_\rho X\qquad \partial_\rho X^1 = 0\qquad \partial_\rho^2 X = V(X) = X \,.
}{eq:2d8}

Solving the constraints \eqref{eq:2d8} establishes boundary conditions for the target space coordinates.
\begin{subequations}
 \label{eq:2d10}
\begin{align}
X &= X_R(\varphi)\,e^\rho + X_L(\varphi)\,e^{-\rho} \label{eq:dilaton} \\
X^0 &= X_R(\varphi)\,e^\rho - X_L(\varphi)\,e^{-\rho}\\ 
X^1 &= X^1(\varphi) 
\end{align}
\end{subequations}
Our result that $X_{R,\, L}$ are allowed to fluctuate in a state-dependent way differs crucially from the approach by Navarro--Salas and Navarro \cite{NavarroSalas:1999up}, who fixed $X_R$ to some (state-independent) constant. 

In addition to the terms that we displayed in the boundary conditions \eqref{eq:2d2}-\eqref{eq:2d10} there could be subleading terms that we are not going to specify explicitly, as they will be of no relevance for our discussion.

The $SL(2)$ Casimir-function turns out to be independent from the radial coordinate $\rho$:
\eq{
C(\varphi) = X^2 - (X^0)^2 - (X^1)^2 = 4 X_R X_L  - (X^1)^2 
}{eq:2d16}
On-shell $C$ is constant and corresponds physically to the mass of the solution. 
The EOM provide three relations between the five state dependent functions, $X^1 T_{R, \, L}=-X^\prime_{R,\, L}$ and $(X^1)^\prime = -2(X_RT_L+X_LT_R)$. 
Above and in what follows we reduce notational clutter by writing $X_{R, \,L}$ and $T_{R,\, L}$ instead of $X_{R, \,L}(\varphi)$ and $T_{R,\, L}(\varphi)$.

As a first check that our boundary conditions above are consistent we show that the action is off-shell gauge-invariant and obeys a well-defined variational principle. Gauge invariance is evident from \eqref{totder}, since in our case the Poisson tensor is linear in the target-space coordinates, so that the term in parentheses in \eqref{totder} vanishes identically. Thus, as long as the variational principle does not require additional boundary terms in the action, off-shell gauge invariance is guaranteed. In this regard our approach differs from our previous treatment \cite{Bergamin:2005pg}, which related the boundary terms in the PSM formulation to the standard boundary terms in the second order formulation, thereby introducing boundary terms that look bizarre from a PSM perspective. Our current approach avoids such boundary terms.

We check now the variational principle.
The first variation of the action \eqref{PSM} yields on-shell
\eq{
\de S\big|_{\textrm{\tiny EOM}} = \frac1\pi\,\int_{\partial M_R}\!\!\!\!\!\!\extd\varphi\,X^I\,\de A_{\varphi\,I}-\frac1\pi\,\int_{\partial M_L}\!\!\!\!\!\!\extd\varphi\,X^I\,\de A_{\varphi\,I}\,.
}{eq:2d43}
Here $\partial M_{R, \,L}$ denote the two disconnected components of the AdS$_2$ boundary, and the relative minus sign appears since the outward pointing unit normals to the boundary have different orientations.
Inserting our boundary and gauge conditions \eqref{eq:2d2}-\eqref{eq:2d10} establishes that all exponentially diverging terms at either boundary vanish identically, and only finite terms remain at each boundary component.
\eq{
\de S\big|_{\textrm{\tiny EOM}} = \frac2\pi\,\Big(\int_{\partial M_R}\!\!\!\!\!\!\extd\varphi-\int_{\partial M_L}\!\!\!\!\!\!\extd\varphi\Big)\,\big(X_R\, \de T_L - X_L\, \de T_R\big) = 0\,.
}{eq:2d44}
Even though the boundary terms do not necessarily vanish individually at each component of the boundary, they cancel each other once both boundary components are taken into account. 

Thus, we have a well-defined variational principle: the first variation of the action vanishes for all variations that preserve our boundary conditions. In this respect our setup differs crucially from the one by Cadoni and Mignemi \cite{Cadoni:1999ja}, who considered similar boundary conditions but only one boundary component, and therefore did not have a well-defined variational principle.

\section{Asymptotic symmetries}

Let us now construct the metric and discuss some of its properties.
With the choices \eqref{eq:2d2}, \eqref{eq:2d3} the line-element is given by
\eq{
\extd s^2 = \extd\rho^2 + \big(T_R^2\,e^{2\rho} + 2T_R T_L + T_L^2\,e^{-2\rho} \big)\,\extd\varphi^2\,.
}{eq:2d4}
Infinitesimal diffeomorphisms $\delta g_{\mu\nu} = \nabla_\mu\xi_\nu + \nabla_\nu\xi_\mu$ that preserve the form \eqref{eq:2d4} of the line-element are generated by vector fields $\xi^\mu$ with
\eq{
\xi^\rho=\xi^\rho(\varphi)\qquad \xi^\varphi=\xi(\varphi) + \frac{\partial_\varphi\xi^\rho\,e^{-\rho}}{2T_R\,e^0_\varphi 
}\,.
}{eq:2d9}
The same vector field also preserves our boundary conditions on the target-space coordinates \eqref{eq:2d10}. In this respect our approach differs crucially from \cite{NavarroSalas:1999up}, where nearly all the diffeomorphisms that preserve their boundary conditions on the metric violate their boundary conditions for the dilaton field.

Defining
$\xi_R = \xi$ and $\xi_L = \xi + (\partial_\varphi\xi^\rho)/(2T_R T_L)$
the state-dependent functions transform as
\begin{align}
 \delta_\xi T_R &= \xi^\rho T_R + \partial_\varphi(\xi_R T_R) \label{eq:2b8} \\
 \delta_\xi T_L &= -\xi^\rho T_L + \partial_\varphi(\xi_L T_L) \label{eq:2b9} \,.
\end{align}
Note that at this stage we have two independent functions of $\varphi$ appearing in the asymptotic symmetries, namely $\xi^\rho(\varphi)$ and $\xi(\varphi)$. This is different from all previous approaches to 2d holography and may provide interesting generalizations.

However, in order to make contact with previous approaches we fix from now on $T_R=\tfrac12$ so that only $T_L$ is allowed to fluctuate. Then $\xi^\rho$ is determined by \eqref{eq:2b8} as $\xi^\rho=-\xi^\prime$ and the transformation law  \eqref{eq:2b9} for $T:=-T_L$ expands to 
\eq{
\delta_\xi T = 2\xi^\prime T + \xi T^\prime + \xi'''\,.
}{eq:2d45} 
What we have just proven is that the state-dependent function $T$ transforms precisely like a chiral stress-tensor in a 2d CFT \cite{diFrancesco} with positive central charge. This is one of our main results.

The PSM gauge transformations \eqref{PSMgauge} that preserve our boundary conditions are generated by gauge parameters $\lambda$ whose components read
\begin{subequations}
 \label{lambdas}
\begin{align}
 \lambda_X &= -\tfrac12 \xi\, e^\rho - \big(T\xi + \xi'' \big)\,e^{-\rho}\\
 \lambda_0 &= \tfrac12 \xi\, e^\rho - \big(T\xi + \xi'' \big)\,e^{-\rho}\\
 \lambda_1 &= -\xi'
\end{align}
\end{subequations}
They are parametrized by a single function $\xi$  and reproduce the transformation law \eqref{eq:2d45}. 
The action of this gauge transformation on the connection 1-forms coincides with a Lie derivative generated by a vector field $\xi$ as given in \eqref{eq:2d9} plus a compensating local Lorentz transformation to maintain our gauge choices.
In addition, the PSM gauge transformations allow to establish transformations laws for all other state-dependent functions. For instance, the transformation law 
\eq{
\de_\xi X_R = X_R \xi' - X_R'\xi 
}{eq:2d38}
shows that $X_R$ behaves like a boundary vector, consistent with the analysis in \cite{Cadoni:2000ah}.

Our main result \eqref{eq:2d45} shows that the central charge is positive, but does not specify its precise value. Without a canonical analysis we do not know how to determine the central charge by direct calculation, and as we shall see in a moment, the canonical analysis is problematic. An indirect way to fix the central charge would be to appeal to the Cardy formula for entropy
\eq{
S_{\textrm{\tiny C}} = 2\pi\,\sqrt{\frac{c h}{6}}
}{eq:cardy}
where $c$ is the central charge and $h$ the value of the Virasoro zero mode charge, 
and to equate it to the Bekenstein--Hawking entropy \cite{Gegenberg:1994pv,Grumiller:2007ju}
\eq{
S_{\textrm{\tiny BH}} = 2\kappa\,X_h
}{eq:SBH}
where $X_h$ is the value of the dilaton field evaluated at the black hole horizon and we have reintroduced the gravitational coupling constant $\kappa$. 
Fixing $X_R=1$ we define 
\eq{
h=\tfrac{\kappa}{4\pi}\,C=\tfrac{2\kappa}{\pi}\,T\,,
}{eq:2d41} 
where $C$ is the Casimir function \eqref{eq:2d16} evaluated on-shell.
We shall provide some justification of the definition \eqref{eq:2d41} below.
Demanding equality between the CFT entropy \eqref{eq:cardy} and the gravitational entropy \eqref{eq:SBH}, $S_{\textrm{\tiny BH}} = 4\ka\,\sqrt{2T}$, yields 
\eq{
2\pi c = 48\kappa\,. 
}{eq:2d40}
It would be nice if there was a direct way to derive this result for the central charge \footnote{
A direct computation by Cadoni and Mignemi led to a discrepancy of $\sqrt{2}$ between CFT and gravitational entropies \cite{Cadoni:1998sg}. Their computation is based upon the assumption that there is a well-defined canonical realization of the asymptotic symmetries, which was contested already in \cite{Park:1999hs}. 
An alternative calculation in \cite{Cadoni:2002rr} yielded $c=6/G$, where $G=1/(8\kappa)$ in our conventions. Their result differs from ours by a factor of $2\pi$, which could be due to the fact that we do not have any $\varphi$-integration in our analysis.
}. 
In fact, a check on the correctness of our results is that the suitably rescaled version of \eqref{eq:2d45}, $\delta_\xi h=2\xi'h+\xi h'+\tfrac{2\kappa}{\pi}\,\xi'''= 2\xi'h+\xi h'+\tfrac{c}{12}\,\xi'''$, with $h$ as defined in \eqref{eq:2d41}, leads precisely to the result \eqref{eq:2d40} for the central charge that we determined from the Cardy formula. Thus, our findings are compatible with the conjecture that 2d dilaton gravity in the linear dilaton vacuum with our boundary conditions is dual to a chiral half of a CFT with central charge $c=24\kappa/\pi$.

We have worked directly with the state-dependent functions and their transformation behavior to unravel the asymptotic symmetries. However, it is fair to ask if the same results, in particular \eqref{eq:2d45}, could have been obtained from a canonical analysis along the lines of \cite{Brown:1986nw}. Surprisingly, the answer is no. Consider the variation of the canonical boundary charges \footnote{%
In Eq.~\eqref{eq:lalapetz} we consider the canonical boundary charges associated with a single AdS$_2$ boundary component. Adding the contributions from both boundary components yields zero.
},
\eq{
\frac{\pi}{\kappa}\,\delta Q = -\lambda_I\,\delta X^I = \xi\,\de X_L + \xi^\prime\,\de X^1 + 2\xi''\,\de X_R + 2\xi T\,\de X_R\,.
}{eq:lalapetz}
The first equality follows by inspection of the derivative terms in the constraints \eqref{eq:2d7} and the second from our boundary conditions and the results \eqref{lambdas}. The only good news is that the charges \eqref{eq:lalapetz} are finite, but they are neither conserved nor integrable. The last term in \eqref{eq:lalapetz} spoils integrability \footnote{
In Ref.~\cite{Cadoni:2000gm} the authors related the Jackiw--Teitelboim model to de~Alfaro--Fubini--Furlan conformal quantum mechanics \cite{deAlfaro:1976je} and also found features resembling non-integrability. Namely, they find an anholonomic constraint, Eq.~(10) in their paper, which leads to a non-exact 1-form. The non-exact expression, translated to our conventions, is $T\extd X_R$, which essentially matches the non-integrable term in our \eqref{eq:lalapetz}.}. 

The non-conservation is well-known and was addressed by Cadoni and Mignemi \cite{Cadoni:1999ja}, who proposed as solution to integrate the ``charges'' over $\varphi$ (which in their conventions is time). However, non-integrability apparently went unnoticed so far and is a serious issue. It implies that there is no good canonical realization of the asymptotic symmetries.

Thus, we are in a similar situation as in flat space holography in four dimensions \cite{Barnich:2011mi,Barnich:2013axa} (see also \cite{Strominger:2013lka,Barnich:2013sxa}): we can consistently define currents and their algebra --- in our case the main result \eqref{eq:2d45} leads to the anomalous transformation law for $T$ familiar from the Virasoro algebra --- but have no conserved integrable canonical charges (see \cite{Wald:1999wa} for a general relativistic discussion). This is a remarkable and unexpected feature of 2d dilaton gravity that deserves further study and could shed light on similar issues arising in flat space holography.

Non-integrability means that we have to pick a certain class of paths in field space to define charges. Let us restrict to field variations that do not change the value of the dilaton field at the right boundary, $\delta X_R=0$. For concreteness we fix $X_R=1$. Then the variation \eqref{eq:lalapetz} simplifies to $\delta Q = \ka/\pi\,\xi\,\de X_L$ which integrates to the charges 
\eq{
Q[\xi] = \tfrac{2k}{\pi}\,\xi\,T = \tfrac{k}{4\pi}\,\xi\,C \,.
}{eq:2d39}
The relation \eqref{eq:2d39} provides the motivation for our definition \eqref{eq:2d41} of the Virasoro zero mode charge. 
(Note that a truncation to $X_R=1$ is consistent only for constant $\xi$.)

\section{Higher spin theories}

We generalize now our results to higher spin theories. Instead of choosing $SL(2)$ we pick some higher rank gauge algebra with generators $L^I$ and structure constants $f^{IJ}{}_K$, i.e., 
$[L^I,L^J]=f^{IJ}{}_K L^K$. Let us fix a representation and assume that the trace form $G^{IJ}={\rm tr}\, (L^IL^J)$ is non-degenrate. Let $G_{JK}$ be the inverse of $G^{IJ}$. 
Matrix-valued fields are defined as $A_\mu\equiv A_{\mu I}L^I$, $X\equiv X^I G_{IJ} L^J$.
Then we define the higher spin theory as a PSM \eqref{PSM} with a linear Poisson tensor \footnote{
Interestingly, also in four dimensions higher spin theory actions can be formulated through higher-dimensional analogs of PSMs \cite{Boulanger:2011dd,Boulanger:2012bj}. 
This is also the appropriate place to mention that PSMs are rigid, in the sense that any Barnich--Henneaux-type of consistent deformation \cite{Barnich:1993vg} of a PSM yields again a PSM with the same field content \cite{Izawa:1999ib}.
} and appropriate identifications of the gauge field 1-forms as zweibein, zuvielbein, and connection.
\eq{
P^{IJ}=f^{IJ}{}_K \,X^K
}{eq:2d49}
The PSM gauge transformations read
\begin{equation}
A_\mu \to e^\lambda\, (\partial_\mu + A_\mu)\,e^{-\lambda} \qquad X\to e^\lambda X e^{-\lambda} \,. \label{gBF}
\end{equation}

For concreteness let us focus on spin-3 gravity, defined by a PSM with $SL(3)$ gauge group, with principally embedded $SL(2)$. The generators are taken as in \cite{Ammon:2012wc}, Eq.~(3.2),
numbered with superscripts, and the indices related to their spin-3 generators $W$ will be taken in parentheses, e.g.~$L^{(-1)}\equiv W_{-1}$. 

In the spin-2 case we have shown that once the boundary conditions on the connection 1-forms are fixed, the boundary conditions for the target space coordinates follow from consistency with the constraints. Therefore, it is sufficient to provide boundary conditions for the connection 1-form $A$. Inspired by the way in which boundary conditions are set up in three-dimensional spin-3 gravity \cite{Henneaux:2010xg,Campoleoni:2010zq} we impose the boundary conditions
\begin{equation}
A = e^{-\rho L_0}\,\big(\extd + a_{\varphi I}L^{I}\,\extd\varphi\big)\,e^{\rho L_0} \label{asusual}
\end{equation}
with
\eq{
a_{\varphi I}L^{I} = L^1 + T(\varphi) L^{-1}+W(\varphi) L^{(-2)}
}{eq:2d50}

The symmetry transformations that preserve the form (\ref{asusual}), \eqref{eq:2d50} depend on two
arbitrary functions, $\lambda_1 = \xi(\varphi)$ and $\lambda_{(2)}=\eta(\varphi)$. 
Decomposing $A_I$ with respect to the $sl(3)$ generators yields the following consistency conditions.
\begin{subequations}
 \label{astr}
\begin{align}
&I=1:& \; &\lambda_0=-\xi' \\
&I=(2):& \; &\lambda_{(1)}=-\eta' \\
&I=0:&\; &\lambda_{-1}=\tfrac 12 (\xi'' +2\xi T -16 \eta W) \\
&I=(1):& \; &\lambda_{(0)}=\tfrac 12 (\eta'' +4\eta T)\\
&I=(0):& \; &\lambda_{(-1)}=\tfrac 13 \big( -\tfrac 12 \eta''' -5\eta'T-2\eta T'\big)\\
&I=(-1):& \; &\lambda_{(-2)}=\tfrac 14 (-\lambda_{(-1)}'+2\lambda_{(0)} T +4 \xi W)
\end{align}
\end{subequations}
The asymptotic symmetry transformations are obtained from (\ref{astr}) by 
$\lambda \to e^{-\rho L^0}\, \lambda \, e^{\rho L^0}$. 

For $I=-1$ and $I=(-2)$ we obtain the transformations of $T$ and $W$, respectively
\begin{eqnarray}
&&\delta T=-\tfrac 12 \xi'''-\xi T'-2T\xi' +12\eta'W+8\eta W'\label{delT}\\
&&\delta W= -\lambda_{(-2)}'+T\lambda_{(-1)} -2W\xi'\,.\label{delW}
\end{eqnarray}
Comparing with \cite{Campoleoni:2010zq} we find that making the identifications $\eta \to \chi$,
$\xi \to - \epsilon$, $T \to 2\pi L/k$ and $W \to \pi W/(2k)$ as well as fixing their $\sigma = -1$ establishes perfect agreement with their equations (4.17)-(4.20). Thus, the current algebra generated by \eqref{delT}, \eqref{delW} is a single copy of the $W_3$ algebra with positive central charge. This is our main result for spin-3 holography.

All the caveats we discussed in the spin-2 case regarding the canonical boundary charges also apply to 2d higher spin theories. In particular, they are non-integrable also for higher spin theories.

We treated here higher-spin theory entirely in its gauge theoretic formulation, since the metric formulation is expected to be more cumbersome. Alkalaev discusses the relation between gauge theoretic and metric formulation in \cite{Alkalaev:2013fsa}.

\section{Discussion and generalizations}

We discussed 2d dilaton gravity with AdS$_2$ boundary conditions in the gauge-theoretic PSM formulation and found as asymptotic symmetry algebra a single copy of the Virasoro algebra, in the sense that we have the anomalous transformation law for the state-dependent function $T$ in \eqref{eq:2d45} with positive central charge. By ``asymptotic symmetry algebra'' we mean all transformations that preserve the gauge- and boundary conditions we imposed, modulo trivial gauge transformations (since we fixed the latter completely we did not need to mod out anything). However, we showed that there is no canonical realization of the asymptotic symmetry algebra due to non-integrability of the canonical boundary ``charges''.

Finally, we formulated generic 2d higher spin theories as PSM with higher rank gauge group and showed that the asymptotic symmetry algebra for spin-3 gravity [with principally embedded $SL(2)$] is a single copy of the $W_3$ algebra. 

We address now generalizations and possible further applications of our results.
Instead of choosing $V=X$ one could study potentials that asymptote to $X$ at large values of $X$, so that curvature asymptotically is still a negative constant \eqref{eq:angelinajolie}. In general such potentials will introduce curvature singularities in the bulk and the global structure no longer is that of global AdS$_2$, i.e., the Penrose diagram no longer is a strip. Even in that case one could still proceed along the lines of our work and take into account both boundaries. 

There are two alternatives. If one wants to consider only a single boundary component then the variational principle is not well-defined for our boundary conditions. One could try to make sense of such a situation, though it is awkward if solutions to the classical EOM no longer are classical solutions of the theory, i.e., they do not extremize the action. It is important to point out that this defect cannot be repaired by adding suitable boundary terms to the action, since the non-vanishing term in the first variation of the on-shell action  \eqref{eq:2d44} is not integrable, for essentially the same reasons that the canonical boundary charges are not integrable \eqref{eq:lalapetz}. It could be possible to find suitable relations between the state-dependent functions $T_{R,\, L}$ and $X_{R, \, L}$ leading  to a well-defined variational principle.

Alternatively, one could fix the leading behavior of the dilaton, $X_R=\rm const.$ in \eqref{eq:2d10}, which reduces the boundary conditions preserving gauge transformations to translations only. Then the theory can no longer be dual to a CFT. Instead, such an approach would generate a correspondence to quantum mechanics at the boundary \cite{Chamon:2011xk}. 

Similar comments apply to generalizations with non-zero kinetic potential, $U(X)\neq 0$ in \eqref{HS}. Also in this case asymptotic AdS$_2$ behavior is guaranteed if the potential $V$ asymptotes to $X$ (times a positive constant).

Adding gauge fields like in \cite{Hartman:2008dq,Castro:prep} basically leads to a modification of the potential $V(X)$, since all gauge fields can be integrated out exactly. For instance, adding to the action \eqref{HS} a Maxwell term $F_{\mu\nu}F^{\mu\nu}$ is equivalent to shifting the potential $V(X)$ by a term that is constant on-shell and proportional to the square of the conserved $U(1)$-charge. Generalizing our boundary conditions to this case is straightforward and only requires to allow for a suitable order unity term in the dilaton in addition to the terms already present in \eqref{eq:dilaton}.

At some point we switched off the fluctuations $T_R$ at one boundary component. It could be interesting to study generalizations where both $T_{R,\, L}$ are switched on. We speculate that they may lead to an additional $u(1)$ current algebra.

Some of the canonical issues of our intrinsic 2d discussion could be avoided by lifting the discussion to higher dimensions. For example, in many approaches to AdS$_2$ holography the starting point is string theory and AdS$_2$ times some compact manifold arises as near horizon approximation to black holes, see for instance \cite{Strominger:1998yg,Sen:2008vm,Castro:prep}. However, it could be rewarding to try to clarify these issues intrinsically within 2d, since this may shed light on more generic situations where an asymptotic symmetry algebra arises, but does not allow for a well-defined canonical realization.

Generalizations to other higher spin theories, like spin-$N$ theories based on $SL(N)$ or Vasiliev-type theories \cite{Vasiliev:1990en} based on hs$(\lambda)$, are straightforward and follow from similar constructions in three dimensions \cite{Henneaux:2010xg,Campoleoni:2010zq}. There will always be a single copy of some $W$-algebra as asymptotic symmetry algebra \footnote{%
      We were informed by Soo-Jong Rey that he formulated higher-spin AdS$_2$ gravity based on
      hs$(\lambda)$ with particular emphasis on relation to higher-spin AdS$_3$ gravity, as reported
      in his talk in 2011 \cite{talk}. He also proposed a specific CFT$_1$ dual based on a variant
      of large-$N$ Calogero model \cite{talk,prep}.}.

\acknowledgments

We are grateful to Arjun Bagchi, Glenn Barnich, Mariano Cadoni, Alejandra Castro, Michael Gary, Radoslav Rashkov, Soo-Jong Rey, Max Riegler and Jan Rosseel for discussions. Additionally, we thank the participants of the workshop ``Higher-Spin and Higher-Curvature Gravity'' in S\~ao Paulo in November 2013 for many fruitful discussions.

DG was supported by the START project Y~435-N16 of the Austrian Science Fund (FWF) and the FWF projects I~952-N16 and I~1030-N27. 
ML was supported by IAFE and University of Buenos Aires.
DV was supported by CNPq and FAPESP.
DG's extended visit to S\~ao Paulo was supported by FAPESP.
DG and ML acknowledge support by a BMWF--MINCyT bilateral cooperation grant, \"OAD project AR 09/2013 and MINCyT project AU/12/09, for their mutual visits.


\begin{thebibliography}{66}
\expandafter\ifx\csname natexlab\endcsname\relax\def\natexlab#1{#1}\fi
\expandafter\ifx\csname bibnamefont\endcsname\relax
  \def\bibnamefont#1{#1}\fi
\expandafter\ifx\csname bibfnamefont\endcsname\relax
  \def\bibfnamefont#1{#1}\fi
\expandafter\ifx\csname citenamefont\endcsname\relax
  \def\citenamefont#1{#1}\fi
\expandafter\ifx\csname url\endcsname\relax
  \def\url#1{\texttt{#1}}\fi
\expandafter\ifx\csname urlprefix\endcsname\relax\def\urlprefix{URL }\fi
\providecommand{\bibinfo}[2]{#2}
\providecommand{\eprint}[2][]{\url{#2}}

\bibitem[{\citenamefont{Mandal et~al.}(1991)\citenamefont{Mandal, Sengupta, and
  Wadia}}]{Mandal:1991tz}
\bibinfo{author}{\bibfnamefont{G.}~\bibnamefont{Mandal}},
  \bibinfo{author}{\bibfnamefont{A.~M.} \bibnamefont{Sengupta}},
  \bibnamefont{and} \bibinfo{author}{\bibfnamefont{S.~R.} \bibnamefont{Wadia}},
  \bibinfo{journal}{Mod. Phys. Lett.} \textbf{\bibinfo{volume}{A6}},
  \bibinfo{pages}{1685} (\bibinfo{year}{1991}).

\bibitem[{\citenamefont{Elitzur et~al.}(1991)\citenamefont{Elitzur, Forge, and
  Rabinovici}}]{Elitzur:1991cb}
\bibinfo{author}{\bibfnamefont{S.}~\bibnamefont{Elitzur}},
  \bibinfo{author}{\bibfnamefont{A.}~\bibnamefont{Forge}}, \bibnamefont{and}
  \bibinfo{author}{\bibfnamefont{E.}~\bibnamefont{Rabinovici}},
  \bibinfo{journal}{Nucl. Phys.} \textbf{\bibinfo{volume}{B359}},
  \bibinfo{pages}{581} (\bibinfo{year}{1991}).

\bibitem[{\citenamefont{Witten}(1991)}]{Witten:1991yr}
\bibinfo{author}{\bibfnamefont{E.}~\bibnamefont{Witten}},
  \bibinfo{journal}{Phys. Rev.} \textbf{\bibinfo{volume}{D44}},
  \bibinfo{pages}{314} (\bibinfo{year}{1991}).

\bibitem[{\citenamefont{Callan et~al.}(1992)\citenamefont{Callan, Giddings,
  Harvey, and Strominger}}]{Callan:1992rs}
\bibinfo{author}{\bibfnamefont{C.~G.} \bibnamefont{Callan},
  \bibfnamefont{Jr.}}, \bibinfo{author}{\bibfnamefont{S.~B.}
  \bibnamefont{Giddings}}, \bibinfo{author}{\bibfnamefont{J.~A.}
  \bibnamefont{Harvey}}, \bibnamefont{and}
  \bibinfo{author}{\bibfnamefont{A.}~\bibnamefont{Strominger}},
  \bibinfo{journal}{Phys. Rev.} \textbf{\bibinfo{volume}{D45}},
  \bibinfo{pages}{1005} (\bibinfo{year}{1992}), \eprint{hep-th/9111056}.

\bibitem[{\citenamefont{Susskind et~al.}(1993)\citenamefont{Susskind,
  Thorlacius, and Uglum}}]{Susskind:1993if}
\bibinfo{author}{\bibfnamefont{L.}~\bibnamefont{Susskind}},
  \bibinfo{author}{\bibfnamefont{L.}~\bibnamefont{Thorlacius}},
  \bibnamefont{and} \bibinfo{author}{\bibfnamefont{J.}~\bibnamefont{Uglum}},
  \bibinfo{journal}{Phys.Rev.} \textbf{\bibinfo{volume}{D48}},
  \bibinfo{pages}{3743} (\bibinfo{year}{1993}), \eprint{hep-th/9306069}.

\bibitem[{\citenamefont{Gegenberg et~al.}(1995)\citenamefont{Gegenberg,
  Kunstatter, and Louis-Martinez}}]{Gegenberg:1994pv}
\bibinfo{author}{\bibfnamefont{J.}~\bibnamefont{Gegenberg}},
  \bibinfo{author}{\bibfnamefont{G.}~\bibnamefont{Kunstatter}},
  \bibnamefont{and}
  \bibinfo{author}{\bibfnamefont{D.}~\bibnamefont{Louis-Martinez}},
  \bibinfo{journal}{Phys. Rev.} \textbf{\bibinfo{volume}{D51}},
  \bibinfo{pages}{1781} (\bibinfo{year}{1995}), \eprint{gr-qc/9408015}.

\bibitem[{\citenamefont{Grumiller and McNees}(2007)}]{Grumiller:2007ju}
\bibinfo{author}{\bibfnamefont{D.}~\bibnamefont{Grumiller}} \bibnamefont{and}
  \bibinfo{author}{\bibfnamefont{R.}~\bibnamefont{McNees}},
  \bibinfo{journal}{JHEP} \textbf{\bibinfo{volume}{04}}, \bibinfo{pages}{074}
  (\bibinfo{year}{2007}), \eprint{hep-th/0703230}.

\bibitem[{\citenamefont{Fiola et~al.}(1994)\citenamefont{Fiola, Preskill,
  Strominger, and Trivedi}}]{Fiola:1994ir}
\bibinfo{author}{\bibfnamefont{T.~M.} \bibnamefont{Fiola}},
  \bibinfo{author}{\bibfnamefont{J.}~\bibnamefont{Preskill}},
  \bibinfo{author}{\bibfnamefont{A.}~\bibnamefont{Strominger}},
  \bibnamefont{and} \bibinfo{author}{\bibfnamefont{S.~P.}
  \bibnamefont{Trivedi}}, \bibinfo{journal}{Phys. Rev.}
  \textbf{\bibinfo{volume}{D50}}, \bibinfo{pages}{3987} (\bibinfo{year}{1994}),
  \eprint{hep-th/9403137}.

\bibitem[{\citenamefont{Ashtekar et~al.}(2008)\citenamefont{Ashtekar, Taveras,
  and Varadarajan}}]{Ashtekar:2008jd}
\bibinfo{author}{\bibfnamefont{A.}~\bibnamefont{Ashtekar}},
  \bibinfo{author}{\bibfnamefont{V.}~\bibnamefont{Taveras}}, \bibnamefont{and}
  \bibinfo{author}{\bibfnamefont{M.}~\bibnamefont{Varadarajan}},
  \bibinfo{journal}{Phys.Rev.Lett.} \textbf{\bibinfo{volume}{100}},
  \bibinfo{pages}{211302} (\bibinfo{year}{2008}), \eprint{0801.1811}.

\bibitem[{\citenamefont{Fischer et~al.}(2001)\citenamefont{Fischer, Grumiller,
  Kummer, and Vassilevich}}]{Fischer:2001vz}
\bibinfo{author}{\bibfnamefont{P.}~\bibnamefont{Fischer}},
  \bibinfo{author}{\bibfnamefont{D.}~\bibnamefont{Grumiller}},
  \bibinfo{author}{\bibfnamefont{W.}~\bibnamefont{Kummer}}, \bibnamefont{and}
  \bibinfo{author}{\bibfnamefont{D.~V.} \bibnamefont{Vassilevich}},
  \bibinfo{journal}{Phys. Lett.} \textbf{\bibinfo{volume}{B521}},
  \bibinfo{pages}{357} (\bibinfo{year}{2001}), \bibinfo{note}{erratum ibid.
  {\bf B532} (2002) 373}, \eprint[http://arXiv.org/abs]{gr-qc/0105034}.

\bibitem[{\citenamefont{Grumiller}(2010)}]{Grumiller:2010bz}
\bibinfo{author}{\bibfnamefont{D.}~\bibnamefont{Grumiller}},
  \bibinfo{journal}{Phys.Rev.Lett.} \textbf{\bibinfo{volume}{105}},
  \bibinfo{pages}{211303} (\bibinfo{year}{2010}), \eprint{1011.3625}.

\bibitem[{\citenamefont{Carlip and Grumiller}(2011)}]{Carlip:2011uc}
\bibinfo{author}{\bibfnamefont{S.}~\bibnamefont{Carlip}} \bibnamefont{and}
  \bibinfo{author}{\bibfnamefont{D.}~\bibnamefont{Grumiller}},
  \bibinfo{journal}{Phys.Rev.} \textbf{\bibinfo{volume}{D84}},
  \bibinfo{pages}{084029} (\bibinfo{year}{2011}), \eprint{1108.4686}.

\bibitem[{\citenamefont{Grumiller et~al.}(2002)\citenamefont{Grumiller, Kummer,
  and Vassilevich}}]{Grumiller:2002nm}
\bibinfo{author}{\bibfnamefont{D.}~\bibnamefont{Grumiller}},
  \bibinfo{author}{\bibfnamefont{W.}~\bibnamefont{Kummer}}, \bibnamefont{and}
  \bibinfo{author}{\bibfnamefont{D.~V.} \bibnamefont{Vassilevich}},
  \bibinfo{journal}{Phys. Rept.} \textbf{\bibinfo{volume}{369}},
  \bibinfo{pages}{327} (\bibinfo{year}{2002}),
  \eprint[http://arXiv.org/abs]{hep-th/0204253}.

\bibitem[{\citenamefont{'t~Hooft}(1993)}]{'tHooft:1993gx}
\bibinfo{author}{\bibfnamefont{G.}~\bibnamefont{'t~Hooft}}, in
  \emph{\bibinfo{booktitle}{Salamfestschrift}} (\bibinfo{publisher}{World
  Scientific}, \bibinfo{year}{1993}), \eprint{gr-qc/9310026}.

\bibitem[{\citenamefont{Susskind}(1995)}]{Susskind:1995vu}
\bibinfo{author}{\bibfnamefont{L.}~\bibnamefont{Susskind}},
  \bibinfo{journal}{J. Math. Phys.} \textbf{\bibinfo{volume}{36}},
  \bibinfo{pages}{6377} (\bibinfo{year}{1995}),
  \eprint[http://arXiv.org/abs]{hep-th/9409089}.

\bibitem[{\citenamefont{Maldacena}(1998)}]{Maldacena:1997re}
\bibinfo{author}{\bibfnamefont{J.~M.} \bibnamefont{Maldacena}},
  \bibinfo{journal}{Adv. Theor. Math. Phys.} \textbf{\bibinfo{volume}{2}},
  \bibinfo{pages}{231} (\bibinfo{year}{1998}), \eprint{hep-th/9711200}.

\bibitem[{\citenamefont{Strominger}(1999)}]{Strominger:1998yg}
\bibinfo{author}{\bibfnamefont{A.}~\bibnamefont{Strominger}},
  \bibinfo{journal}{JHEP} \textbf{\bibinfo{volume}{01}}, \bibinfo{pages}{007}
  (\bibinfo{year}{1999}), \eprint{hep-th/9809027}.

\bibitem[{\citenamefont{Cadoni and
  Mignemi}(1999{\natexlab{a}})}]{Cadoni:1999ja}
\bibinfo{author}{\bibfnamefont{M.}~\bibnamefont{Cadoni}} \bibnamefont{and}
  \bibinfo{author}{\bibfnamefont{S.}~\bibnamefont{Mignemi}},
  \bibinfo{journal}{Nucl. Phys.} \textbf{\bibinfo{volume}{B557}},
  \bibinfo{pages}{165} (\bibinfo{year}{1999}{\natexlab{a}}),
  \eprint{hep-th/9902040}.

\bibitem[{\citenamefont{Navarro-Salas and Navarro}(2000)}]{NavarroSalas:1999up}
\bibinfo{author}{\bibfnamefont{J.}~\bibnamefont{Navarro-Salas}}
  \bibnamefont{and} \bibinfo{author}{\bibfnamefont{P.}~\bibnamefont{Navarro}},
  \bibinfo{journal}{Nucl. Phys.} \textbf{\bibinfo{volume}{B579}},
  \bibinfo{pages}{250} (\bibinfo{year}{2000}), \eprint{hep-th/9910076}.

\bibitem[{\citenamefont{Hartman and Strominger}(2009)}]{Hartman:2008dq}
\bibinfo{author}{\bibfnamefont{T.}~\bibnamefont{Hartman}} \bibnamefont{and}
  \bibinfo{author}{\bibfnamefont{A.}~\bibnamefont{Strominger}},
  \bibinfo{journal}{JHEP} \textbf{\bibinfo{volume}{0904}}, \bibinfo{pages}{026}
  (\bibinfo{year}{2009}), \eprint{0803.3621}.

\bibitem[{\citenamefont{Henneaux and Rey}(2010)}]{Henneaux:2010xg}
\bibinfo{author}{\bibfnamefont{M.}~\bibnamefont{Henneaux}} \bibnamefont{and}
  \bibinfo{author}{\bibfnamefont{S.-J.} \bibnamefont{Rey}},
  \bibinfo{journal}{JHEP} \textbf{\bibinfo{volume}{1012}}, \bibinfo{pages}{007}
  (\bibinfo{year}{2010}), \eprint{1008.4579}.

\bibitem[{\citenamefont{Campoleoni et~al.}(2010)\citenamefont{Campoleoni,
  Fredenhagen, Pfenninger, and Theisen}}]{Campoleoni:2010zq}
\bibinfo{author}{\bibfnamefont{A.}~\bibnamefont{Campoleoni}},
  \bibinfo{author}{\bibfnamefont{S.}~\bibnamefont{Fredenhagen}},
  \bibinfo{author}{\bibfnamefont{S.}~\bibnamefont{Pfenninger}},
  \bibnamefont{and} \bibinfo{author}{\bibfnamefont{S.}~\bibnamefont{Theisen}},
  \bibinfo{journal}{JHEP} \textbf{\bibinfo{volume}{1011}}, \bibinfo{pages}{007}
  (\bibinfo{year}{2010}), \eprint{1008.4744}.

\bibitem[{\citenamefont{Gaberdiel and Gopakumar}(2011)}]{Gaberdiel:2010pz}
\bibinfo{author}{\bibfnamefont{M.~R.} \bibnamefont{Gaberdiel}}
  \bibnamefont{and}
  \bibinfo{author}{\bibfnamefont{R.}~\bibnamefont{Gopakumar}},
  \bibinfo{journal}{Phys.Rev.} \textbf{\bibinfo{volume}{D83}},
  \bibinfo{pages}{066007} (\bibinfo{year}{2011}), \eprint{1011.2986}.

\bibitem[{\citenamefont{Afshar et~al.}(2013)\citenamefont{Afshar, Bagchi,
  Fareghbal, Grumiller, and Rosseel}}]{Afshar:2013vka}
\bibinfo{author}{\bibfnamefont{H.}~\bibnamefont{Afshar}},
  \bibinfo{author}{\bibfnamefont{A.}~\bibnamefont{Bagchi}},
  \bibinfo{author}{\bibfnamefont{R.}~\bibnamefont{Fareghbal}},
  \bibinfo{author}{\bibfnamefont{D.}~\bibnamefont{Grumiller}},
  \bibnamefont{and} \bibinfo{author}{\bibfnamefont{J.}~\bibnamefont{Rosseel}},
  \bibinfo{journal}{Phys.Rev.Lett.} \textbf{\bibinfo{volume}{111}},
  \bibinfo{pages}{121603} (\bibinfo{year}{2013}), \eprint{1307.4768}.

\bibitem[{\citenamefont{Gonzalez et~al.}(2013)\citenamefont{Gonzalez, Matulich,
  Pino, and Troncoso}}]{Gonzalez:2013oaa}
\bibinfo{author}{\bibfnamefont{H.~A.} \bibnamefont{Gonzalez}},
  \bibinfo{author}{\bibfnamefont{J.}~\bibnamefont{Matulich}},
  \bibinfo{author}{\bibfnamefont{M.}~\bibnamefont{Pino}}, \bibnamefont{and}
  \bibinfo{author}{\bibfnamefont{R.}~\bibnamefont{Troncoso}},
  \bibinfo{journal}{JHEP} \textbf{\bibinfo{volume}{1309}}, \bibinfo{pages}{016}
  (\bibinfo{year}{2013}), \eprint{1307.5651}.

\bibitem[{\citenamefont{Banks and O'Loughlin}(1991)}]{Banks:1991mk}
\bibinfo{author}{\bibfnamefont{T.}~\bibnamefont{Banks}} \bibnamefont{and}
  \bibinfo{author}{\bibfnamefont{M.}~\bibnamefont{O'Loughlin}},
  \bibinfo{journal}{Nucl. Phys.} \textbf{\bibinfo{volume}{B362}},
  \bibinfo{pages}{649} (\bibinfo{year}{1991}).

\bibitem[{\citenamefont{Russo and Tseytlin}(1992)}]{Russo:1992yg}
\bibinfo{author}{\bibfnamefont{J.~G.} \bibnamefont{Russo}} \bibnamefont{and}
  \bibinfo{author}{\bibfnamefont{A.~A.} \bibnamefont{Tseytlin}},
  \bibinfo{journal}{Nucl. Phys.} \textbf{\bibinfo{volume}{B382}},
  \bibinfo{pages}{259} (\bibinfo{year}{1992}), \eprint{arXiv:hep-th/9201021}.

\bibitem[{\citenamefont{Odintsov and Shapiro}(1991)}]{Odintsov:1991qu}
\bibinfo{author}{\bibfnamefont{S.~D.} \bibnamefont{Odintsov}} \bibnamefont{and}
  \bibinfo{author}{\bibfnamefont{I.~L.} \bibnamefont{Shapiro}},
  \bibinfo{journal}{Phys. Lett.} \textbf{\bibinfo{volume}{B263}},
  \bibinfo{pages}{183} (\bibinfo{year}{1991}).

\bibitem[{\citenamefont{Kummer et~al.}(1997)\citenamefont{Kummer, Liebl, and
  Vassilevich}}]{Kummer:1996hy}
\bibinfo{author}{\bibfnamefont{W.}~\bibnamefont{Kummer}},
  \bibinfo{author}{\bibfnamefont{H.}~\bibnamefont{Liebl}}, \bibnamefont{and}
  \bibinfo{author}{\bibfnamefont{D.~V.} \bibnamefont{Vassilevich}},
  \bibinfo{journal}{Nucl. Phys.} \textbf{\bibinfo{volume}{B493}},
  \bibinfo{pages}{491} (\bibinfo{year}{1997}),
  \eprint[http://arXiv.org/abs]{gr-qc/9612012}.

\bibitem[{\citenamefont{Brown and Henneaux}(1986)}]{Brown:1986nw}
\bibinfo{author}{\bibfnamefont{J.~D.} \bibnamefont{Brown}} \bibnamefont{and}
  \bibinfo{author}{\bibfnamefont{M.}~\bibnamefont{Henneaux}},
  \bibinfo{journal}{Commun. Math. Phys.} \textbf{\bibinfo{volume}{104}},
  \bibinfo{pages}{207} (\bibinfo{year}{1986}).

\bibitem[{\citenamefont{Bagchi et~al.}(2012)\citenamefont{Bagchi, Detournay,
  and Grumiller}}]{Bagchi:2012yk}
\bibinfo{author}{\bibfnamefont{A.}~\bibnamefont{Bagchi}},
  \bibinfo{author}{\bibfnamefont{S.}~\bibnamefont{Detournay}},
  \bibnamefont{and}
  \bibinfo{author}{\bibfnamefont{D.}~\bibnamefont{Grumiller}},
  \bibinfo{journal}{Phys.Rev.Lett.} \textbf{\bibinfo{volume}{109}},
  \bibinfo{pages}{151301} (\bibinfo{year}{2012}), \eprint{1208.1658}.

\bibitem[{\citenamefont{Achucarro and Townsend}(1986)}]{Achucarro:1986vz}
\bibinfo{author}{\bibfnamefont{A.}~\bibnamefont{Achucarro}} \bibnamefont{and}
  \bibinfo{author}{\bibfnamefont{P.~K.} \bibnamefont{Townsend}},
  \bibinfo{journal}{Phys. Lett.} \textbf{\bibinfo{volume}{B180}},
  \bibinfo{pages}{89} (\bibinfo{year}{1986}).

\bibitem[{\citenamefont{Witten}(1988)}]{Witten:1988hc}
\bibinfo{author}{\bibfnamefont{E.}~\bibnamefont{Witten}},
  \bibinfo{journal}{Nucl. Phys.} \textbf{\bibinfo{volume}{B311}},
  \bibinfo{pages}{46} (\bibinfo{year}{1988}).

\bibitem[{\citenamefont{Ikeda}(1994)}]{Ikeda:1993fh}
\bibinfo{author}{\bibfnamefont{N.}~\bibnamefont{Ikeda}},
  \bibinfo{journal}{Annals Phys.} \textbf{\bibinfo{volume}{235}},
  \bibinfo{pages}{435} (\bibinfo{year}{1994}), \eprint{hep-th/9312059}.

\bibitem[{\citenamefont{Schaller and Strobl}(1994)}]{Schaller:1994es}
\bibinfo{author}{\bibfnamefont{P.}~\bibnamefont{Schaller}} \bibnamefont{and}
  \bibinfo{author}{\bibfnamefont{T.}~\bibnamefont{Strobl}},
  \bibinfo{journal}{Mod. Phys. Lett.} \textbf{\bibinfo{volume}{A9}},
  \bibinfo{pages}{3129} (\bibinfo{year}{1994}),
  \eprint[http://arXiv.org/abs]{hep-th/9405110}.

\bibitem[{\citenamefont{Strobl}(2004)}]{Strobl:2004im}
\bibinfo{author}{\bibfnamefont{T.}~\bibnamefont{Strobl}},
  \bibinfo{journal}{Phys.Rev.Lett.} \textbf{\bibinfo{volume}{93}},
  \bibinfo{pages}{211601} (\bibinfo{year}{2004}), \eprint{hep-th/0406215}.

\bibitem[{\citenamefont{Castro and Song}()}]{Castro:prep}
\bibinfo{author}{\bibfnamefont{A.}~\bibnamefont{Castro}} \bibnamefont{and}
  \bibinfo{author}{\bibfnamefont{W.}~\bibnamefont{Song}}, \bibinfo{note}{in
  preparation}.

\bibitem[{\citenamefont{Alkalaev}(2013)}]{Alkalaev:2013fsa}
\bibinfo{author}{\bibfnamefont{K.}~\bibnamefont{Alkalaev}}
  (\bibinfo{year}{2013}), \eprint{1311.5119}.

\bibitem[{\citenamefont{Jackiw}(1984)}]{Jackiw:1984}
\bibinfo{author}{\bibfnamefont{R.}~\bibnamefont{Jackiw}}, in
  \emph{\bibinfo{booktitle}{Quantum Theory Of Gravity}}, edited by
  \bibinfo{editor}{\bibfnamefont{S.}~\bibnamefont{Christensen}}
  (\bibinfo{publisher}{Adam Hilger}, \bibinfo{address}{Bristol},
  \bibinfo{year}{1984}), pp. \bibinfo{pages}{403--420}.

\bibitem[{\citenamefont{Teitelboim}(1984)}]{Teitelboim:1984}
\bibinfo{author}{\bibfnamefont{C.}~\bibnamefont{Teitelboim}}, in
  \emph{\bibinfo{booktitle}{Quantum Theory Of Gravity}}, edited by
  \bibinfo{editor}{\bibfnamefont{S.}~\bibnamefont{Christensen}}
  (\bibinfo{publisher}{Adam Hilger}, \bibinfo{address}{Bristol},
  \bibinfo{year}{1984}), pp. \bibinfo{pages}{327--344}.

\bibitem[{\citenamefont{Isler and Trugenberger}(1989)}]{Isler:1989hq}
\bibinfo{author}{\bibfnamefont{K.}~\bibnamefont{Isler}} \bibnamefont{and}
  \bibinfo{author}{\bibfnamefont{C.~A.} \bibnamefont{Trugenberger}},
  \bibinfo{journal}{Phys. Rev. Lett.} \textbf{\bibinfo{volume}{63}},
  \bibinfo{pages}{834} (\bibinfo{year}{1989}).

\bibitem[{\citenamefont{Chamseddine and Wyler}(1989)}]{Chamseddine:1989yz}
\bibinfo{author}{\bibfnamefont{A.~H.} \bibnamefont{Chamseddine}}
  \bibnamefont{and} \bibinfo{author}{\bibfnamefont{D.}~\bibnamefont{Wyler}},
  \bibinfo{journal}{Phys. Lett.} \textbf{\bibinfo{volume}{B228}},
  \bibinfo{pages}{75} (\bibinfo{year}{1989}).

\bibitem[{\citenamefont{Cangemi and Jackiw}(1992)}]{Cangemi:1992bj}
\bibinfo{author}{\bibfnamefont{D.}~\bibnamefont{Cangemi}} \bibnamefont{and}
  \bibinfo{author}{\bibfnamefont{R.}~\bibnamefont{Jackiw}},
  \bibinfo{journal}{Phys. Rev. Lett.} \textbf{\bibinfo{volume}{69}},
  \bibinfo{pages}{233} (\bibinfo{year}{1992}),
  \eprint[http://arXiv.org/abs]{hep-th/9203056}.

\bibitem[{\citenamefont{Achucarro}(1993)}]{Achucarro:1992mb}
\bibinfo{author}{\bibfnamefont{A.}~\bibnamefont{Achucarro}},
  \bibinfo{journal}{Phys. Rev. Lett.} \textbf{\bibinfo{volume}{70}},
  \bibinfo{pages}{1037} (\bibinfo{year}{1993}), \eprint{hep-th/9207108}.

\bibitem[{\citenamefont{Vassilevich}(2013)}]{Vassilevich:2013ai}
\bibinfo{author}{\bibfnamefont{D.}~\bibnamefont{Vassilevich}},
  \bibinfo{journal}{Phys.Rev.} \textbf{\bibinfo{volume}{D87}},
  \bibinfo{pages}{104011} (\bibinfo{year}{2013}), \eprint{1301.7029}.

\bibitem[{\citenamefont{Bergamin et~al.}(2006)\citenamefont{Bergamin,
  Grumiller, Kummer, and Vassilevich}}]{Bergamin:2005pg}
\bibinfo{author}{\bibfnamefont{L.}~\bibnamefont{Bergamin}},
  \bibinfo{author}{\bibfnamefont{D.}~\bibnamefont{Grumiller}},
  \bibinfo{author}{\bibfnamefont{W.}~\bibnamefont{Kummer}}, \bibnamefont{and}
  \bibinfo{author}{\bibfnamefont{D.~V.} \bibnamefont{Vassilevich}},
  \bibinfo{journal}{Class. Quant. Grav.} \textbf{\bibinfo{volume}{23}},
  \bibinfo{pages}{3075} (\bibinfo{year}{2006}), \eprint{hep-th/0512230}.

\bibitem[{\citenamefont{Di~Francesco et~al.}(1997)\citenamefont{Di~Francesco,
  Mathieu, and Senechal}}]{diFrancesco}
\bibinfo{author}{\bibfnamefont{P.}~\bibnamefont{Di~Francesco}},
  \bibinfo{author}{\bibfnamefont{P.}~\bibnamefont{Mathieu}}, \bibnamefont{and}
  \bibinfo{author}{\bibfnamefont{D.}~\bibnamefont{Senechal}},
  \emph{\bibinfo{title}{Conformal Field Theory}}
  (\bibinfo{publisher}{Springer}, \bibinfo{year}{1997}).

\bibitem[{\citenamefont{Cadoni and Mignemi}(2000)}]{Cadoni:2000ah}
\bibinfo{author}{\bibfnamefont{M.}~\bibnamefont{Cadoni}} \bibnamefont{and}
  \bibinfo{author}{\bibfnamefont{S.}~\bibnamefont{Mignemi}},
  \bibinfo{journal}{Phys.Lett.} \textbf{\bibinfo{volume}{B490}},
  \bibinfo{pages}{131} (\bibinfo{year}{2000}), \eprint{hep-th/0002256}.

\bibitem[{\citenamefont{Barnich and Troessaert}(2011)}]{Barnich:2011mi}
\bibinfo{author}{\bibfnamefont{G.}~\bibnamefont{Barnich}} \bibnamefont{and}
  \bibinfo{author}{\bibfnamefont{C.}~\bibnamefont{Troessaert}},
  \bibinfo{journal}{JHEP} \textbf{\bibinfo{volume}{1112}}, \bibinfo{pages}{105}
  (\bibinfo{year}{2011}), \eprint{1106.0213}.

\bibitem[{\citenamefont{Barnich and Troessaert}(2013)}]{Barnich:2013axa}
\bibinfo{author}{\bibfnamefont{G.}~\bibnamefont{Barnich}} \bibnamefont{and}
  \bibinfo{author}{\bibfnamefont{C.}~\bibnamefont{Troessaert}},
  \bibinfo{journal}{JHEP} \textbf{\bibinfo{volume}{1311}}, \bibinfo{pages}{003}
  (\bibinfo{year}{2013}), \eprint{1309.0794}.

\bibitem[{\citenamefont{Strominger}(2013)}]{Strominger:2013lka}
\bibinfo{author}{\bibfnamefont{A.}~\bibnamefont{Strominger}}
  (\bibinfo{year}{2013}), \eprint{1308.0589}.

\bibitem[{\citenamefont{Barnich and Lambert}(2013)}]{Barnich:2013sxa}
\bibinfo{author}{\bibfnamefont{G.}~\bibnamefont{Barnich}} \bibnamefont{and}
  \bibinfo{author}{\bibfnamefont{P.-H.} \bibnamefont{Lambert}}
  (\bibinfo{year}{2013}), \eprint{1310.2698}.

\bibitem[{\citenamefont{Wald and Zoupas}(2000)}]{Wald:1999wa}
\bibinfo{author}{\bibfnamefont{R.~M.} \bibnamefont{Wald}} \bibnamefont{and}
  \bibinfo{author}{\bibfnamefont{A.}~\bibnamefont{Zoupas}},
  \bibinfo{journal}{Phys.Rev.} \textbf{\bibinfo{volume}{D61}},
  \bibinfo{pages}{084027} (\bibinfo{year}{2000}), \eprint{gr-qc/9911095}.

\bibitem[{\citenamefont{Ammon et~al.}(2013)\citenamefont{Ammon, Gutperle,
  Kraus, and Perlmutter}}]{Ammon:2012wc}
\bibinfo{author}{\bibfnamefont{M.}~\bibnamefont{Ammon}},
  \bibinfo{author}{\bibfnamefont{M.}~\bibnamefont{Gutperle}},
  \bibinfo{author}{\bibfnamefont{P.}~\bibnamefont{Kraus}}, \bibnamefont{and}
  \bibinfo{author}{\bibfnamefont{E.}~\bibnamefont{Perlmutter}},
  \bibinfo{journal}{J.Phys.} \textbf{\bibinfo{volume}{A46}},
  \bibinfo{pages}{214001} (\bibinfo{year}{2013}), \eprint{1208.5182}.

\bibitem[{\citenamefont{Chamon et~al.}(2011)\citenamefont{Chamon, Jackiw, Pi,
  and Santos}}]{Chamon:2011xk}
\bibinfo{author}{\bibfnamefont{C.}~\bibnamefont{Chamon}},
  \bibinfo{author}{\bibfnamefont{R.}~\bibnamefont{Jackiw}},
  \bibinfo{author}{\bibfnamefont{S.-Y.} \bibnamefont{Pi}}, \bibnamefont{and}
  \bibinfo{author}{\bibfnamefont{L.}~\bibnamefont{Santos}},
  \bibinfo{journal}{Phys.Lett.} \textbf{\bibinfo{volume}{B701}},
  \bibinfo{pages}{503} (\bibinfo{year}{2011}), \eprint{1106.0726}.

\bibitem[{\citenamefont{Sen}(2009)}]{Sen:2008vm}
\bibinfo{author}{\bibfnamefont{A.}~\bibnamefont{Sen}},
  \bibinfo{journal}{Int.J.Mod.Phys.} \textbf{\bibinfo{volume}{A24}},
  \bibinfo{pages}{4225} (\bibinfo{year}{2009}), \eprint{0809.3304}.

\bibitem[{\citenamefont{Vasiliev}(1990)}]{Vasiliev:1990en}
\bibinfo{author}{\bibfnamefont{M.~A.} \bibnamefont{Vasiliev}},
  \bibinfo{journal}{Phys.Lett.} \textbf{\bibinfo{volume}{B243}},
  \bibinfo{pages}{378} (\bibinfo{year}{1990}).

\bibitem[{\citenamefont{Cadoni and
  Mignemi}(1999{\natexlab{b}})}]{Cadoni:1998sg}
\bibinfo{author}{\bibfnamefont{M.}~\bibnamefont{Cadoni}} \bibnamefont{and}
  \bibinfo{author}{\bibfnamefont{S.}~\bibnamefont{Mignemi}},
  \bibinfo{journal}{Phys. Rev.} \textbf{\bibinfo{volume}{D59}},
  \bibinfo{pages}{081501} (\bibinfo{year}{1999}{\natexlab{b}}),
  \eprint{hep-th/9810251}.

\bibitem[{\citenamefont{Park and Yee}(2000)}]{Park:1999hs}
\bibinfo{author}{\bibfnamefont{M.-I.} \bibnamefont{Park}} \bibnamefont{and}
  \bibinfo{author}{\bibfnamefont{J.~H.} \bibnamefont{Yee}},
  \bibinfo{journal}{Phys. Rev.} \textbf{\bibinfo{volume}{D61}},
  \bibinfo{pages}{088501} (\bibinfo{year}{2000}), \eprint{hep-th/9910213}.

\bibitem[{\citenamefont{Cadoni et~al.}(2002)\citenamefont{Cadoni, Carta, and
  Mignemi}}]{Cadoni:2002rr}
\bibinfo{author}{\bibfnamefont{M.}~\bibnamefont{Cadoni}},
  \bibinfo{author}{\bibfnamefont{P.}~\bibnamefont{Carta}}, \bibnamefont{and}
  \bibinfo{author}{\bibfnamefont{S.}~\bibnamefont{Mignemi}},
  \bibinfo{journal}{Nucl.Phys.} \textbf{\bibinfo{volume}{B632}},
  \bibinfo{pages}{383} (\bibinfo{year}{2002}), \eprint{hep-th/0202180}.

\bibitem[{\citenamefont{Cadoni et~al.}(2001)\citenamefont{Cadoni, Carta, Klemm,
  and Mignemi}}]{Cadoni:2000gm}
\bibinfo{author}{\bibfnamefont{M.}~\bibnamefont{Cadoni}},
  \bibinfo{author}{\bibfnamefont{P.}~\bibnamefont{Carta}},
  \bibinfo{author}{\bibfnamefont{D.}~\bibnamefont{Klemm}}, \bibnamefont{and}
  \bibinfo{author}{\bibfnamefont{S.}~\bibnamefont{Mignemi}},
  \bibinfo{journal}{Phys.Rev.} \textbf{\bibinfo{volume}{D63}},
  \bibinfo{pages}{125021} (\bibinfo{year}{2001}), \eprint{hep-th/0009185}.

\bibitem[{\citenamefont{de~Alfaro et~al.}(1976)\citenamefont{de~Alfaro, Fubini,
  and Furlan}}]{deAlfaro:1976je}
\bibinfo{author}{\bibfnamefont{V.}~\bibnamefont{de~Alfaro}},
  \bibinfo{author}{\bibfnamefont{S.}~\bibnamefont{Fubini}}, \bibnamefont{and}
  \bibinfo{author}{\bibfnamefont{G.}~\bibnamefont{Furlan}},
  \bibinfo{journal}{Nuovo Cim.} \textbf{\bibinfo{volume}{A34}},
  \bibinfo{pages}{569} (\bibinfo{year}{1976}).

\bibitem[{\citenamefont{Boulanger and Sundell}(2011)}]{Boulanger:2011dd}
\bibinfo{author}{\bibfnamefont{N.}~\bibnamefont{Boulanger}} \bibnamefont{and}
  \bibinfo{author}{\bibfnamefont{P.}~\bibnamefont{Sundell}},
  \bibinfo{journal}{J.Phys.} \textbf{\bibinfo{volume}{A44}},
  \bibinfo{pages}{495402} (\bibinfo{year}{2011}), \eprint{1102.2219}.

\bibitem[{\citenamefont{Boulanger et~al.}(2012)\citenamefont{Boulanger,
  Colombo, and Sundell}}]{Boulanger:2012bj}
\bibinfo{author}{\bibfnamefont{N.}~\bibnamefont{Boulanger}},
  \bibinfo{author}{\bibfnamefont{N.}~\bibnamefont{Colombo}}, \bibnamefont{and}
  \bibinfo{author}{\bibfnamefont{P.}~\bibnamefont{Sundell}},
  \bibinfo{journal}{JHEP} \textbf{\bibinfo{volume}{1210}}, \bibinfo{pages}{043}
  (\bibinfo{year}{2012}), \eprint{1205.3339}.

\bibitem[{\citenamefont{Barnich and Henneaux}(1993)}]{Barnich:1993vg}
\bibinfo{author}{\bibfnamefont{G.}~\bibnamefont{Barnich}} \bibnamefont{and}
  \bibinfo{author}{\bibfnamefont{M.}~\bibnamefont{Henneaux}},
  \bibinfo{journal}{Phys. Lett.} \textbf{\bibinfo{volume}{B311}},
  \bibinfo{pages}{123} (\bibinfo{year}{1993}),
  \eprint[http://arXiv.org/abs]{hep-th/9304057}.

\bibitem[{\citenamefont{Izawa}(2000)}]{Izawa:1999ib}
\bibinfo{author}{\bibfnamefont{K.~I.} \bibnamefont{Izawa}},
  \bibinfo{journal}{Prog. Theor. Phys.} \textbf{\bibinfo{volume}{103}},
  \bibinfo{pages}{225} (\bibinfo{year}{2000}),
  \eprint[http://arXiv.org/abs]{hep-th/9910133}.

\bibitem[{\citenamefont{Rey}(2011)}]{talk}
\bibinfo{author}{\bibfnamefont{S.-J.} \bibnamefont{Rey}}, \bibinfo{note}{talk at the Simons Workshop on Higher Spin Gravity and Holography}, \bibinfo{year}{March 2011}, \href{http://media.scgp.stonybrook.edu/video/video.php?f=20110316_3_qtp.mp4}{link to video of the talk}.

\bibitem[{\citenamefont{Rey}(2014)}]{prep}
\bibinfo{author}{\bibfnamefont{S.-J.} \bibnamefont{Rey}}, \bibinfo{note}{personal communication and in preparation}.

\end{thebibliography}

\end{document}